\documentclass{moriond}
\usepackage{color,graphicx,epsfig}
\usepackage{ifpdf}
\usepackage{amsmath}
\usepackage{bm}
\usepackage{color}
\usepackage[english]{babel}
\usepackage{graphicx}
\usepackage{amsfonts}
\usepackage{amssymb}
\usepackage{braket}
\usepackage{hyperref}
\usepackage[utf8]{inputenc}

\definecolor{nicered}{rgb}{0.7,0.1,0.1}
\definecolor{nicegreen}{rgb}{0.1,0.5,0.1}
\hypersetup{colorlinks,citecolor= nicegreen,linkcolor= nicered}

\newcommand{\kc}{\kappa_c}
\newcommand{\kb}{\kappa_b}

\newcommand{\kq}{\kappa_Q}

\newcommand{\gluon}{gg \to hj}
\newcommand{\quark}{gQ \to hQ,\, Q\bar{Q}\to hg}

\newcommand{\Photo}{}

\begin{document}
\title{Yukawas and trilinear Higgs terms from loops}

\author{Ulrich Haisch}

\address{Rudolf Peierls Centre for Theoretical Physics,
   University of Oxford, \\ OX1 3NP Oxford, United Kingdom}

\maketitle\abstracts{
It is illustrated how LHC precision measurements of rates and distributions in single-Higgs production can be used to constrain the charm Yukawa as well as the Higgs trilinear coupling.}

\section{Setting the stage}

The interactions of the standard model (SM) Higgs boson are determined by the following Lagrangian density 
\begin{equation} \label{Eq1}
{\cal L} \supset \left | D_\mu H \right |^2  - \sum_f \left ( y_f \bar f_L H f_R + {\rm h.c.} \right ) - V \,, \qquad 
V = -\mu \left | H \right |^2 + \lambda \left | H \right |^4 \,,
\end{equation}
where $D_\mu$ is the $SU(2)_L \times U(1)_Y$ covariant derivative, $H$ is the Higgs doublet, the subscripts~$L,R$ denote the chirality of fermionic fields and $y_f$ are the corresponding Yukawa couplings. 

What do we know about the above interactions? From the ATLAS and CMS combination of the~LHC~Run~I measurements of the Higgs boson production and decay rates,\hspace{0.5mm}\cite{Khachatryan:2016vau} it follows that the gauge-Higgs interactions, as encoded in the term~$\left | D_\mu H \right |^2$, are at the level of ${\cal O} (10\%)$~SM-like. The Yukawa interactions~$y_f \bar f_L H f_R + {\rm h.c.}$, on the other hand, have been tested with this accuracy only in the case of the tau lepton, while the constraints on the top and bottom Yukawa couplings just reach the~${\cal O} (20\%)$ level. Apart from the muon Yukawa coupling which is marginally constrained by the combined ATLAS and CMS analysis, first and second generation Yukawa couplings are not directly probed at present.  In the case of the Higgs potential $V$, we know the vacuum expectation value (VEV) of $H$ for a long time, and the discovery of a spin-0 CP-even state of  $m_h \simeq 125 \, {\rm GeV}$ at the LHC tells us about the second derivative of $V$ around its VEV, as  this quantity determines the Higgs mass. The trilinear and quartic Higgs self-interactions that result from (\ref{Eq1}) are however essentially untested at the moment. 

In the following it will be shown that LHC precision measurements of rates and distributions in single-Higgs production can be used to constrain some of the presently poorly known Higgs interactions terms appearing in (\ref{Eq1}). The two explicit examples that we will discuss in some detail are the  charm Yukawa coupling  and the Higgs trilinear coupling.

\section{Charm Yukawa coupling}

It has been common lore\hspace{0.75mm}\cite{Peskin:2012we} that extractions of $y_c$ can only be performed with a few-percent uncertainty at an $e^+e^-$ machine such as the ILC.\hspace{0.5mm}\cite{Ono:2013sea} Gaining direct access to $y_c$ is however not hopeless, since in its high-luminosity run the LHC (HL-LHC) will produce around $1.7 \cdot 10^8$ Higgses bosons per experiment with~$3 \, {\rm ab}^{-1}$ of integrated luminosity.\hspace{0.5mm}\cite{Dawson:2013bba} In fact, several different strategies have been proposed to constrain modifications $\kc = y_c/y_c^{\rm SM}$. A first way to probe~$\kc$ consists in searching for the exclusive decay~$h \to J/\psi \gamma$.\hspace{0.5mm}\cite{Bodwin:2013gca,Kagan:2014ila,Koenig:2015pha} While reconstructing the~$J/\psi$ via its dimuon decay leads to a clean experimental signature, the small branching ratio of $1.8 \cdot 10^{-7}$, implies that only 30 signal events can be expected at  each ATLAS and~CMS. This makes a detection challenging given the large continuous background due to~QCD production of charmonia and a jet faking a photon.\hspace{0.5mm}\cite{Aad:2015sda,Perez:2015lra} The process $h\to c \bar c \gamma$ can also be used to bound $\kappa_c$ and the constraining power has recently been found to be at least comparable to that of $h \to J/\psi \gamma$.\hspace{0.5mm}\cite{Han:2017yhy} Strategies with larger signal cross sections are $p p \to V  c \bar c$ where $V = W, Z$,\hspace{0.5mm}\cite{Perez:2015lra,Perez:2015aoa} the $p p \to h c$\hspace{0.75mm}\cite{Brivio:2015fxa} channel and $gg\to h\to c\bar{c}$.\hspace{0.5mm}\cite{Delaunay:2013pja} These searches rely on charm tagging~($c$-tagging). Since $c$-tagging algorithms at ATLAS and CMS are currently inefficient, bottom jets cannot be discriminated perfectly from charm jets so that the latter modes  not only measure $\kc$, but certain linear combinations of~$\kc$ and~$\kb = y_b/y_b^{\rm SM}$. Notice that despite its lower acceptance in pseudorapidity, LHCb has recently also obtained a first limit on~$pp \to V c \bar c$,\hspace{0.5mm}\cite{LHCb-CONF-2016-006} and hence in the long run   might be able to set relevant bounds on the modification~$\kappa_c$ as well. 

Another independent procedure to constrain $\kappa_c$,\hspace{0.75mm}\cite{Bishara:2016jga}  that does not suffer one of the aforementioned limitations, is based on the observation that the cross section in gluon-fusion  Higgs production provides sensitivity to  $\kappa_t =  y_t/y_t^{\rm SM}$ and $\kappa_b$ through the interference of top and bottom loops\hspace{0.75mm}\cite{Khachatryan:2016vau} 
\begin{equation} \label{Eq2}
\sigma \left ( g g \to h \right )  \propto 1.06 \hspace{0.25mm} \kappa_t^2 + 0.01 \hspace{0.25mm} \kappa_b^2 - 0.07  \hspace{0.25mm} \kappa_t \kappa_b \,.
\end{equation}
Such interference effects appear not only in the total rate, but in all $gg \to h j$ distributions such as the transverse momentum $p_{T,h}$ of the Higgs boson. In fact, these contributions are dynamically enhanced by  logarithms\hspace{0.75mm}\cite{Baur:1989cm} of the form $\kappa_Q \hspace{0.25mm} m_Q^2/m_h^2 \ln^2 \big ( p_{T,h}^2/m_Q^2 \big )$ with $Q = b,c$. If instead the Higgs is produced in $\quark$, the resulting leading order~(LO) differential cross section scales as $\kq^2$, with an additional suppression factor of ${\cal O} (\alpha_s/\pi)$ for each initial-state sea-quark parton distribution function which is generated from gluon splitting. Due to the different Lorentz structure of the amplitudes in the $m_Q\to 0$ limit, the $\gluon$ and $\quark$ processes do not interfere at ${\cal O} (\alpha_s^2)$. This ensures that no terms scaling linearly in~$\kq$ are present in the $\quark$ channels at this order. 

\begin{figure}[!t]
\begin{center}
\includegraphics[width=0.4625\columnwidth]{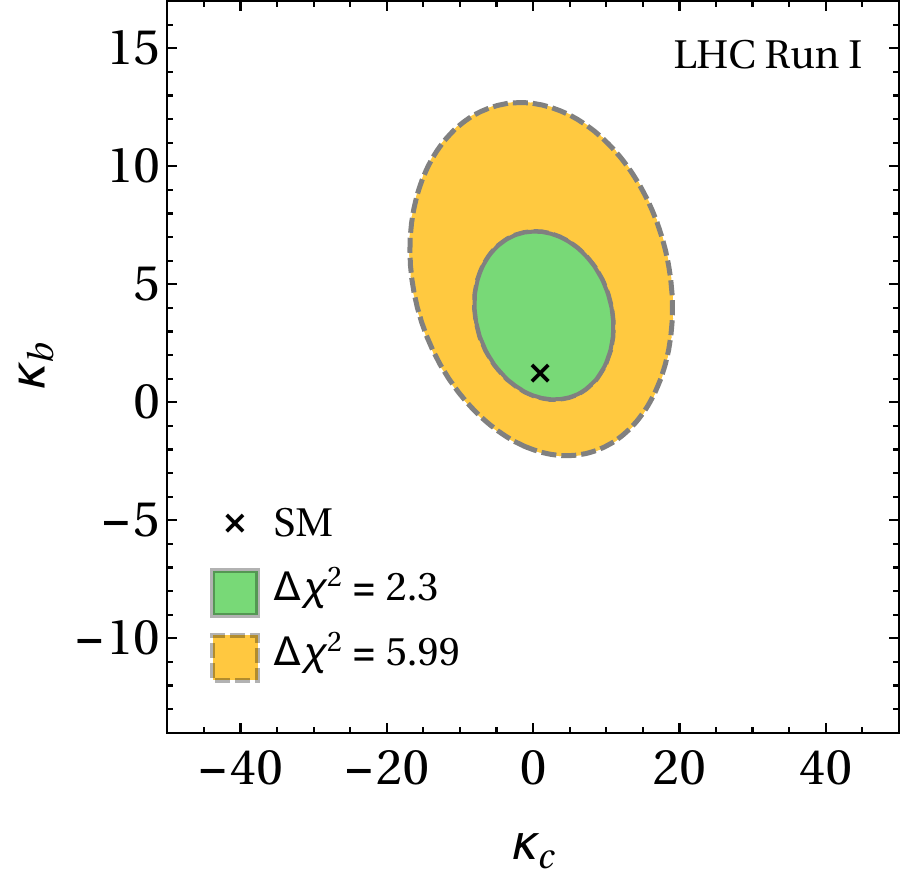}  \qquad 
\includegraphics[width=0.45\columnwidth]{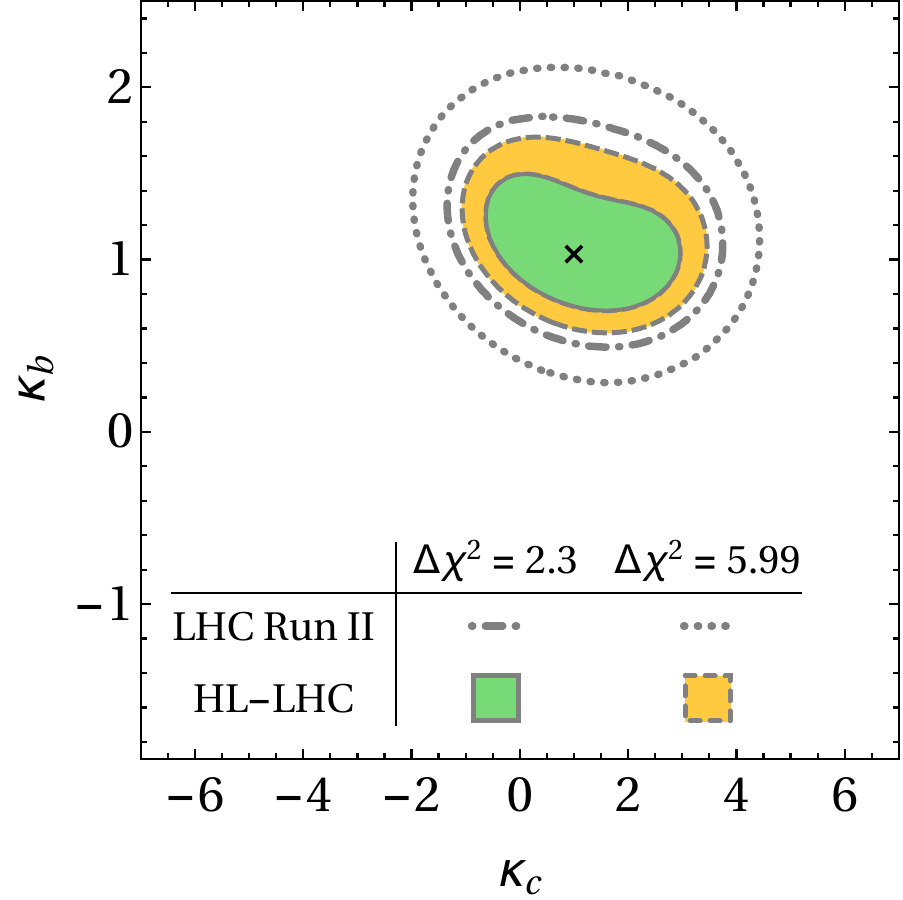}  
\vspace{0mm}
\caption{\label{Fig1} Left: The $\Delta\chi^2=2.3$  and $\Delta\chi^2=5.99$ regions in the $\kc \hspace{0.25mm}$--$\hspace{0.25mm} \kb$ plane following from the combination of the  ATLAS measurements of the normalised $p_{T,h}$ distribution in  the $h \to \gamma \gamma$ and $h \to Z Z^\ast \to 4 \ell$ channels. The SM point is indicated by the black cross. Right: Projected future constraints in the $\kc \hspace{0.25mm}$--$\hspace{0.25mm} \kb$ plane.  The figure shows  projections for the LHC Run II (HL-LHC) with $0.3 \, {\rm ab}^{-1}$ ($3 \, {\rm ab}^{-1}$) of integrated luminosity at $\sqrt{s} = 13 \, {\rm TeV}$.}
\end{center}
\end{figure}

Since  the gluon-fusion and quark-initiated processes lead to different $p_{T,h}$  distributions, the two Higgs production mechanism can be experimentally disentangled. This feature has been exploited to set constraints on $y_{b,c,s}$\hspace{0.75mm}\cite{Bishara:2016jga} as well as $y_{u,d}$.\hspace{0.5mm}\cite{Soreq:2016rae} In particular, it has been shown that in the case of the bottom and charm Yukawa couplings both the effects linear and quadratic in $\kappa_{b,c}$ can be phenomenologically relevant, while in the case of the light quarks only terms proportional to~$\kappa_{s,u,d}^2$ matter. Since the deviations of the $p_{T,h}$ spectrum amount to only several percent for~${\cal O} (1)$ modifications of $\kappa_{b,c}$~$\big ($as expected from (\ref{Eq2})$\big )$, precise theoretical predictions for the $\gluon$ and $\quark$ channels are needed to derive faithful bounds on the bottom and charm Yukawa couplings. In the case of $\gluon$, next-to-leading order~(NLO) corrections to the spectrum in the Higgs effective field theory (HEFT)\hspace{0.75mm}\cite{deFlorian:1999zd,Ravindran:2002dc,Glosser:2002gm} are included using~\texttt{MCFM}.\hspace{0.5mm}\cite{Campbell:2015qma} The total cross sections for inclusive Higgs production are obtained from \texttt{HIGLU},\hspace{0.5mm}\cite{Spira:1995mt} taking into account the next-to-next-to-leading order~(NNLO) corrections in the~HEFT.\hspace{0.5mm}\cite{Harlander:2002wh,Anastasiou:2002yz,Ravindran:2003um} Sudakov logarithms of the form $\ln \left (p_{T,h}/m_h \right)$ are resummed up to next-to-next-to-leading logarithmic order.\hspace{0.75mm}\cite{Bozzi:2003jy,Becher:2010tm,Monni:2016ktx} The $\quark$ contributions to the $p_{T,h}$ distribution are calculated at NLO with \texttt{MG5$_{ }$aMC@NLO}.\hspace{0.5mm}\cite{Wiesemann:2014ioa} The theoretical uncertainties  obtained in this way amount to around $\pm5\%$\hspace{0.75mm}\cite{Bishara:2016jga} and could be  improved by taking into account recent theoretical developments in $\gluon$\hspace{0.75mm}\cite{Becher:2013xia,Melnikov:2016emg,Lindert:2017pky} and $\quark$.\hspace{0.5mm}\cite{Harlander:2014hya}  Since  non-perturbative corrections to the $p_{T,h}$ distribution are not larger than $\pm2\%$ in the region of moderate $p_{T,h}$,\hspace{0.5mm}\cite{Bishara:2016jga,Becher:2012yn} the theoretical predictions can be expected to reach  an accuracy of a few percent in the not too far future. 

\begin{table}[!t]
\begin{center}
\begin{tabular}{|c|c|c|c|}
\hline
Method & LHC Run I  &   LHC Run II & HL-LHC   \\ \hline \hline
$h \to J/\psi \gamma$\hspace{1mm}\cite{Koenig:2015pha,Perez:2015lra} & $|\kappa_c| <  429$ & $|\kappa_c| \lesssim 80$ & $|\kappa_c| \lesssim 45$ \\ \hline
$h \to c \bar c \gamma$\hspace{1mm}\cite{Han:2017yhy} & --- &  --- & $|\kappa_c| < 6.3$ \\ \hline
$pp \to V c \bar c$\hspace{1.25mm}\cite{Perez:2015lra} & $|\kappa_c| <  234$ & $|\kappa_c| < 21$ & $|\kappa_c| < 3.7$ \\ \hline
$pp \to h c$\hspace{1mm}\cite{Brivio:2015fxa} & --- & --- & $|\kappa_c| < 2.6$ \\ \hline
$p_{T,h}$ spectrum\hspace{1mm}\cite{Bishara:2016jga} & $\kappa_c \in [-16,18]$ & $\kappa_c \in [-1.4, 3.8]$ & $\kappa_c \in [-0.6,3.0]$ \\ \hline
\end{tabular}
\vspace{0mm}
\caption{\label{Tab1} Sensitivities for probing the modification $\kappa_c$ of the charm Yukawa coupling with various methods. The~95\%~CL bounds as quoted in the literature after LHC Run I and II as well as the HL-LHC phase are given. }
\end{center}
\end{table}

In order to derive the current constraints on~$\kb$ and~$\kc$, we harness the normalised~$p_{T,h}$ distribution in inclusive Higgs production.\hspace{0.5mm}\cite{Aad:2015lha} This spectrum is obtained by ATLAS from a combination of $h \to \gamma \gamma$ and $h \to Z Z^\ast \to 4 \ell$ decays and based on $20.3 \, {\rm fb}^{-1}$ of $\sqrt{s} = 8 \, {\rm TeV}$ data. In our analysis, we include the first seven bins in the range $p_{T,h} \in [0,100]$\,GeV whose experimental uncertainty is dominated by the statistical error. In the left panel of Figure~\ref{Fig1} the $\Delta\chi^2=2.3$  and $\Delta\chi^2=5.99$ contours (corresponding to a $68\%$ and $95\%$ confidence level~(CL) for a Gaussian distribution) in the $\kc \hspace{0.25mm}$--$\hspace{0.25mm} \kb$ plane are displayed. By profiling over $\kb$, one obtains the following~$95\%$~CL bound on $\kc$\hspace{0.75mm}\cite{Bishara:2016jga} 
\begin{equation} \label{Eq3} 
\kc \in [-16,18]  \,, \qquad (\text{LHC Run I}) \,.  
\end{equation}
As can be seen from Table~\ref{Tab1}, this  limit is significantly stronger than the existing bounds on the charm Yukawa coupling from $h \to J/\psi \gamma$ and $pp \to V c \bar c$. It is also more stringent than the  limit $|\kc| \lesssim 130$ following from the measurements of the total Higgs width, but it is not competitive with the bound~$|\kc| \lesssim 6.2$ that derives from a global analysis of LHC Run I Higgs data.\hspace{0.75mm}\cite{Perez:2015aoa}

We study two benchmark cases to demonstrate the LHC prospects of extracting $\kappa_c$ through analyses of the $p_{T,h}$ spectrum. Our LHC Run II scenario employs~$0.3 \, {\rm ab}^{-1}$ of integrated luminosity and assumes a systematic error of~$\pm3\%$ on the experimental side and a total theoretical uncertainty of $\pm5\%$. This means that we envision that the non-statistical uncertainties present at LHC Run~I can be halved in the coming years, which seems plausible. Our HL-LHC scenario instead uses~$3 \, {\rm ab}^{-1}$ of data and foresees a reduction of both systematic and theoretical errors by another factor of two, leading to uncertainties of~$\pm1.5\%$ and~$\pm2.5\%$, respectively.  We stress that this last scenario is illustrative of the reach that can be achieved with improved theory uncertainties. The corresponding fit results are presented on the right-hand side in Figure~\ref{Fig1}. The unshaded contours refer to the LHC Run II scenario with the dot-dashed (dotted) lines corresponding to $\Delta\chi^2=2.3\;(5.99)$. Analogously, the shaded contours with the solid (dashed) lines refer to the HL-LHC. By profiling over $\kb$, one finds in the LHC~Run~II scenario the following~$95\%$~CL bound on the $y_c$ modifications\hspace{0.75mm}\cite{Bishara:2016jga} 
\begin{equation} \label{Eq4}
\kc \in [-1.4,3.8] \,, \qquad  (\text{LHC Run II})\,,
\end{equation}
while the corresponding HL-LHC bound reads\hspace{0.75mm}\cite{Bishara:2016jga} 
\begin{equation} \label{Eq5}
\kc \in [-0.6,3.0]  \,, \qquad  (\text{HL-LHC})\,.
\end{equation}
As is evident from Table~\ref{Tab1},  these limits compare well not only with the projected reach of other proposed strategies but also have the nice feature that they are controlled by the systematic uncertainties that can be reached in the future. This is not the case for extractions of $y_c$ using the $h \to J/\psi \gamma$, $h \to c \bar c \gamma$, $p p \to V c \bar c$ and $p p \to h c$ channels, which are either limited by small signal-to-background ratios or by the charm-bottom discrimination of heavy-flavour tagging.

\section{Trilinear Higgs coupling}

After electroweak (EW) symmetry breaking the  self-interactions of the Higgs field $h$ in the SM can be parameterised  by 
\begin{equation} \label{Eq6}
V \supset \lambda v h^3 + \frac{\chi}{4} \hspace{0.25mm} h^4 \,, \qquad \lambda = \chi = \frac{m_h^2}{2v^2} \,,
\end{equation}
One way to experimentally constrain the coefficients $\lambda$ and $\chi$ consists in measuring double-Higgs and triple-Higgs production. Since the cross section for $pp \to 3h$ production is of ${\cal O} (0.1\, {\rm fb})$ at $\sqrt{s} = 14 \, {\rm TeV}$ even the HL-LHC will only allow to set very loose bounds on the Higgs quartic. The prospects to observe double-Higgs production at the HL-LHC is considerably better because the $pp \to hh$ cross section amounts to ${\cal O} (33 \, {\rm fb})$ at the same centre-of-mass energy. Measuring double-Higgs production at the HL-LHC however still remains challenging  and as a result even with the full data set of $3 \, {\rm ab}^{-1}$ only an ${\cal O} (1)$ determination of the trilinear Higgs coupling $\lambda$ seems possible under optimistic assumptions.

The coefficient $\lambda$ is however also subject to indirect  constraints from  processes such as single-Higgs production\hspace{0.75mm}\cite{McCullough:2013rea,Gorbahn:2016uoy,Degrassi:2016wml,Bizon:2016wgr} or EW precision observables\hspace{0.75mm}\cite{Degrassi:2017ucl,Kribs:2017znd} since a modified $h^3$ coupling alters these observables at the loop level. In order to describe modifications of the trilinear Higgs coupling in a model-independent fashion, one can employ the SM effective field theory and add dimension-six operators to the SM Lagrangian density
\begin{equation} \label{Eq7}
{\cal L}^{(6)} = \sum_k \frac{\bar c_k}{v^2} \hspace{0.5mm} O_k \,, \qquad O_6 = - \lambda \left |H \right |^6 \,,
\end{equation}
where $v \simeq 246 \,{\rm GeV}$ denotes the Higgs VEV. Under the assumption that the operator $O_6$ represents the only relevant modification of the Higgs self-interactions at tree level, instead of the result (\ref{Eq6}) one then finds 
\begin{equation} \label{Eq8}
V \supset \kappa_\lambda  \hspace{0.25mm} \lambda v h^3 + \kappa_\chi \hspace{0.25mm} \frac{\chi}{4} \hspace{0.25mm}  h^4 \,, \qquad 
\kappa_\lambda = 1 + \bar c_6 \,,  \qquad 
\kappa_\chi = 1 + 6 \hspace{0.25mm} \bar c_6 \,.
\end{equation}
These relations allow one to parameterise a modified trilinear Higgs coupling via the Wilson coefficient $\bar c_6 = \kappa_\lambda -1$ or equivalent $\kappa_\lambda$. Other operators such as  $O_H = \big (\partial_\mu |H|^2 \big )^2$ or $O_8 =  \left |H \right |^8$ also change the $h^3$ coupling at tree level, but will not be discussed in what follows.

The operator $O_6$ introduced in (\ref{Eq7}) modifies vector boson fusion (VBF), associated $Vh$\hspace{0.75mm}\cite{Degrassi:2016wml,Bizon:2016wgr}  as well as $t \bar t h$ production\hspace{0.75mm}\cite{Degrassi:2016wml} at  the one-loop level, while it enters the gluon-fusion channel at two loops.\hspace{0.5mm}\cite{Gorbahn:2016uoy,Degrassi:2016wml} Higgs decays to fermions, $W$ and $Z$ pairs are altered at one loop,\hspace{0.75mm}\cite{Degrassi:2016wml,Bizon:2016wgr} while modifications of the digluon and diphoton rates are  again a two-loop effect.\hspace{0.5mm}\cite{Gorbahn:2016uoy,Degrassi:2016wml} All production and decay channels receive two types of contributions: firstly, a process dependent one, which is linear in~$\bar c_6$ and secondly, a universal one associated to the Higgs wave function renormalisation, which contains a piece quadratic in $\bar c_6$. In order to give an impression of the complexity of the corresponding perturbative calculations, let us quote an explicit expression for the non-universal part of the two-loop $gg \to h$ form factor. Performing an asymptotic expansion in the ratio $r = m_h^2/m_t^2$, one finds that the sought contribution is proportional to\hspace{0.75mm}\cite{GHinprep}    
\begin{eqnarray} \label{Eq9}
\begin{split}
& \ln r + \frac{\pi}{\sqrt3} - \frac{23}{12}  + 
r \left ( \frac{7}{10} \hspace{0.5mm} \ln r + \frac{7 \pi}{20 \sqrt{3}} - \frac{259}{240} \right ) +
r^2 \left ( \frac{349}{1008} \hspace{0.5mm} \ln r + \frac{23 \pi}{240 \sqrt{3}} - \frac{464419}{1058400} \right ) \\[2mm] 
& + r^3 \left ( \frac{1741}{10800} \hspace{0.5mm} \ln r + \frac{13 \pi}{525 \sqrt{3}} -  \frac{31795373}{190512000} \right ) 
+ r^4 \left ( \frac{10817}{138600} \hspace{0.5mm} \ln r + \frac{1789 \pi}{277200 \sqrt{3}} -  \frac{40370773}{614718720} \right ) \\[2mm] 
& + r^5 \left ( \frac{2798759}{68796000} \hspace{0.5mm} \ln r + \frac{439357 \pi}{252252000 \sqrt{3}} -  \frac{2551088981767}{90901530720000} \right )  + {\cal O} (r^6) \,,   
\end{split}
\end{eqnarray}
where the terms up to order $r^3$ have already been given before,\hspace{0.5mm}\cite{Degrassi:2016wml} while the $r^4$ and $r^5$ terms are presented here for the first time. 

So how do the direct and indirect  limits on $\kappa_\lambda$ compare after LHC Run I, if only the trilinear Higgs coupling is allowed to deviate from the SM? Performing a $\chi^2$ fit with $\Delta \chi^2 = 3.84$ corresponding to a $95\%$~CL for a Gaussian distribution, one obtains from double-Higgs production\hspace{0.75mm}\cite{Gorbahn:2016uoy}
\begin{equation} \label{Eq10}
\kappa_\lambda \in [-14.5, 19.1] \,, \qquad \text{($pp \to hh$ at LHC Run I)} \,, 
\end{equation}
while the combination of the LHC Run~I single-Higgs data\hspace{0.75mm}\cite{Khachatryan:2016vau}  leads to 
\begin{equation} \label{Eq11}
\kappa_\lambda \in [-7.7, 15.1] \,, \qquad \text{($pp \to h$ at LHC Run I)}  \,.
\end{equation}
The quoted limit from $pp \to h$ compares well with other existing single-Higgs extractions\hspace{0.75mm}\cite{Gorbahn:2016uoy,Degrassi:2016wml,Bizon:2016wgr} and is slightly more stringent than (\ref{Eq10}) as well as the bound that can be derived from EW precision observables.\hspace{0.5mm}\cite{Degrassi:2017ucl,Kribs:2017znd} It has been derived by combining the results for LO gluon-fusion\hspace{0.75mm}\cite{Degrassi:2016wml,GHinprep} and  NNLO VBF and $Vh$ production\hspace{0.75mm}\cite{Bizon:2016wgr}  with that of LO $t \bar t h$ production.\hspace{0.75mm}\cite{Degrassi:2016wml}

The results  (\ref{Eq10}) and  (\ref{Eq11}) indicate that  to exploit the full LHC potential all available  informations on the $h^3$ term should be combined. Most of the existing studies of indirect constraints on the trilinear Higgs coupling are based on the simplified assumption that only the $h^3$ vertex is modified while all other Higgs interactions remain SM-like. Recently\hspace{0.75mm}\cite{DiVita:2017eyz} this assumption has been dropped and ten parameter fits  allowing for modifications $\kappa_\lambda$ have been performed. In this way it has been shown that standard global Higgs analyses suffer from degeneracies that prevent one from extracting robust bounds on each individual coupling (or Wilson coefficient) once large non-standard $h^3$ interactions are considered. The inclusion of $pp \to hh$ production as well as the use of differential measurements in the associated single-Higgs production channels $Vh$ and $t \bar t h$, can however help to overcome the limitations of a global Higgs-coupling fit. Including differential information on both single-Higgs and double-Higgs production, one finds from the ten parameter fit the following $95\%$~CL limit\hspace{0.75mm}\cite{DiVita:2017eyz}
\begin{equation} \label{Eq12}
\kappa_\lambda \in [-0.7, 7.1] \,, \qquad \text{($pp \to h$ and $pp \to hh$ differential at HL-LHC)} \,,
\end{equation}
assuming an integrated luminosity of $3 \, {\rm ab}^{-1}$. To which extent the result (\ref{Eq12}) represents the ultimate limit on $\kappa_\lambda$ that can be obtained at the HL-LHC requires further study, in particular a  detailed assessment of the experimental uncertainties entering the global $\chi^2$ analysis. 

In order to further illustrate the importance to measure differential Higgs distributions and to include them into global analyses of Higgs couplings, we consider besides (\ref{Eq7}) the following three dimension-six operators 
\begin{equation} \label{Eq13}
\begin{split}
O_{HW} &= \frac{8i}{g} \hspace{0.5mm} \big ( D_\mu H^\dagger \tau^i D_\nu H \big ) W^{i, \mu \nu} \,, \\[2mm]
O_{W} &= \frac{4i}{g}  \hspace{0.5mm} \big  (  H^\dagger \tau^i \! \stackrel{\leftrightarrow}{D}_\mu \! H \big ) D_\nu W^{i, \mu \nu} \,, \\[2mm] 
O_{B} &= \frac{2i g^\prime}{g}  \hspace{0.5mm} \big (  H^\dagger \! \stackrel{\leftrightarrow}{D}_\mu \! H \big ) D_\nu B^{\mu \nu} \,,
\end{split}
\end{equation}
which unlike $O_6$ modify the $VVh$ vertex at tree level. In (\ref{Eq13}) the variables $g$ and $g^\prime$ denote the $SU(2)_L$ and $U(1)_Y$ gauge coupling, respectively, $W^{i,\mu \nu}$ and $B^{\mu \nu}$ are the corresponding field-strength tensors, the derivative operator $\stackrel{\leftrightarrow}{D}_\mu$ is defined as $H^\dagger \! \stackrel{\leftrightarrow}{D}_\mu \! H = H^\dagger D_\mu H - \big ( D_\mu H^\dagger \big) H$ and $\tau^i = \sigma^i/2$ with $\sigma^i$ the usual Pauli matrices.

\begin{figure}[!t]
\begin{center}
\includegraphics[width=0.45\columnwidth]{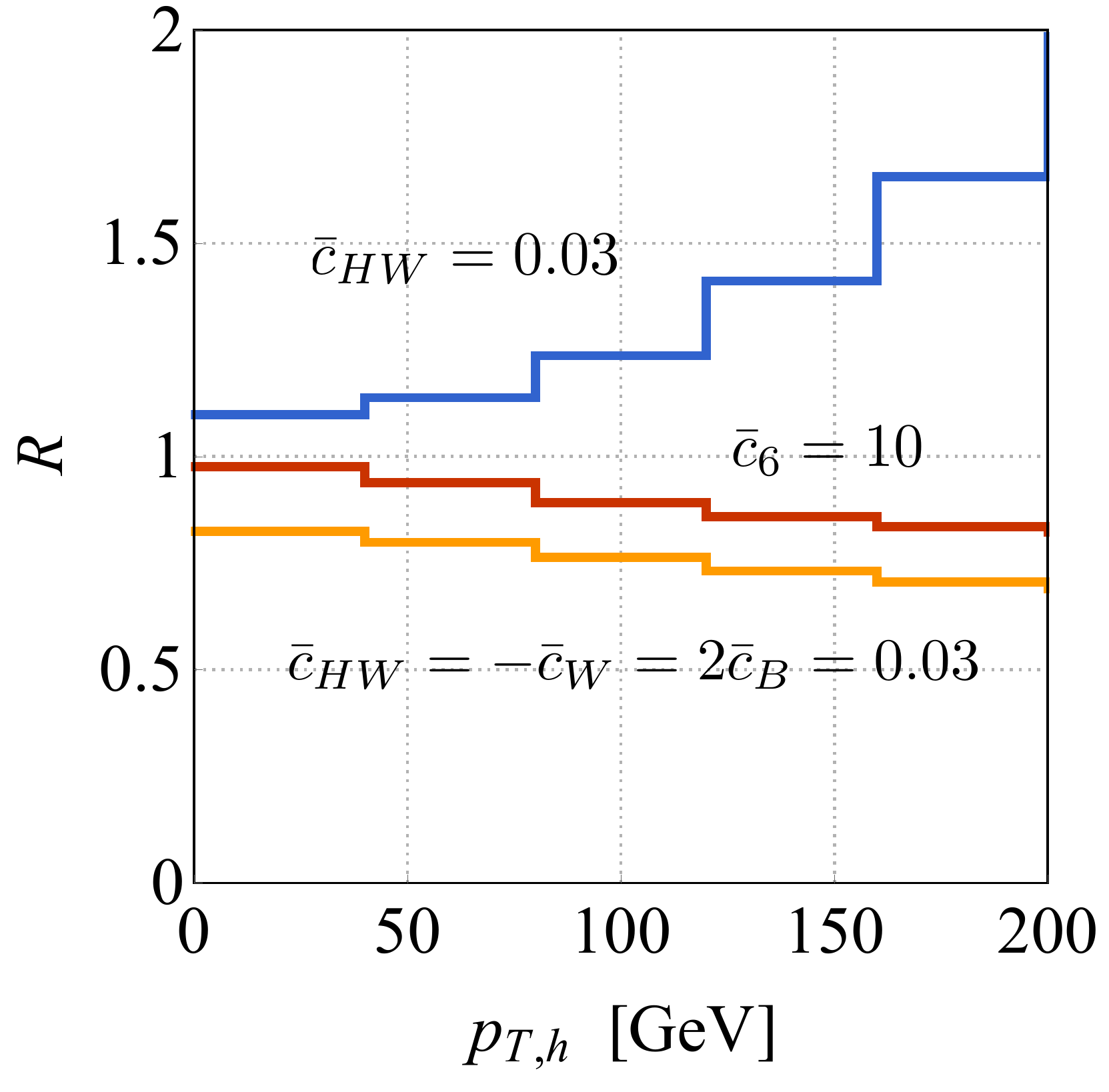}  \qquad 
\includegraphics[width=0.45\columnwidth]{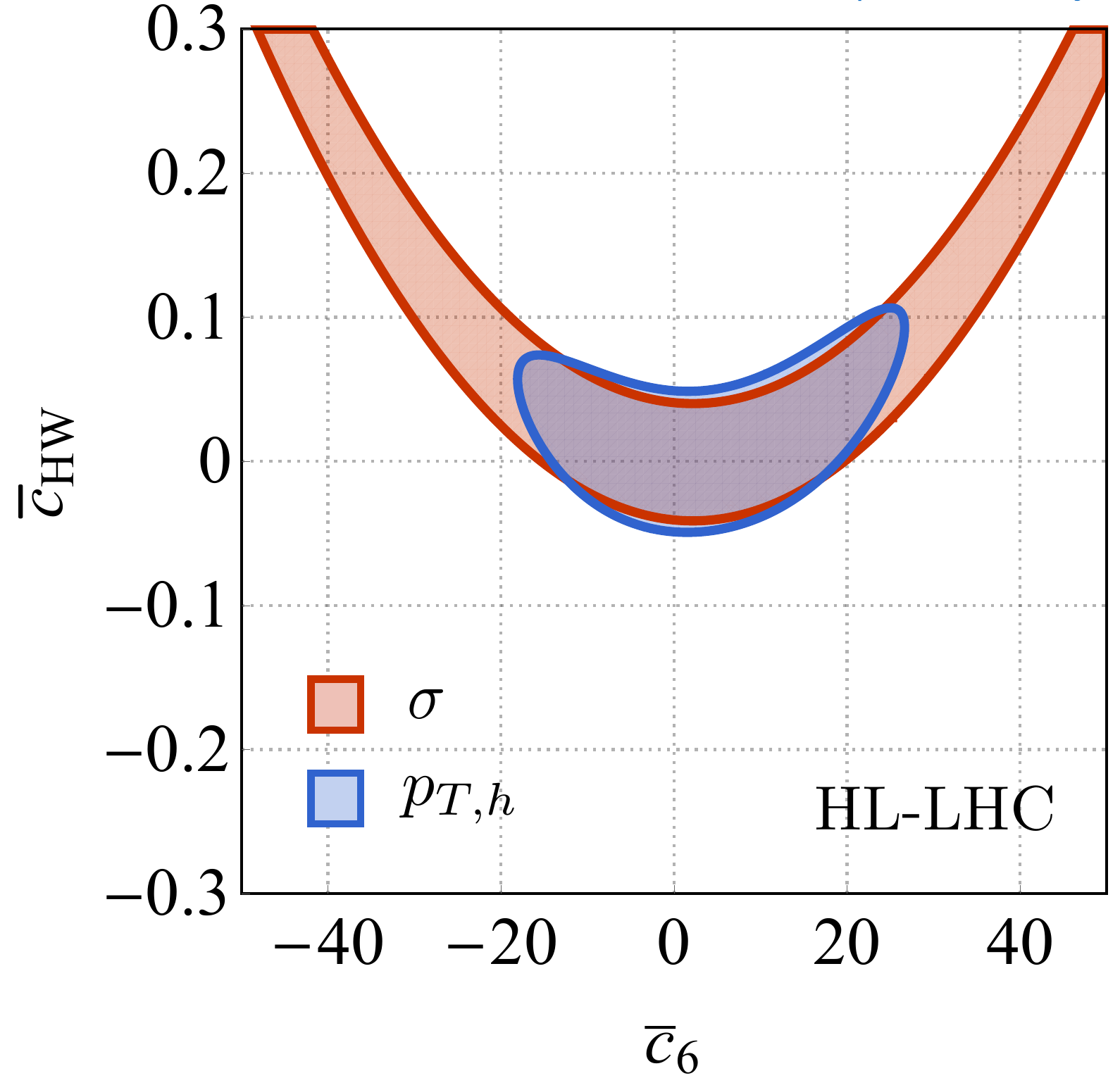}  
\vspace{0mm}
\caption{\label{Fig2} Left: The $p_{T,h}$ distributions in $Wh$ production normalised to the SM spectrum at $\sqrt{s} = 13 \, {\rm TeV}$. The coloured curves correspond to different choices of Wilson coefficients. The coefficients $\bar c_k$ not indicated in the plot are set to zero. Right: Constraints at $95\%$~CL in the $\bar c_6\hspace{0.5mm}$--$\hspace{0.5mm} \bar c_{HW}$ plane that  follow from a hypothetical measurement of the $Wh$ channel at the HL-LHC. The red contour is obtained by a fit to the inclusive cross section, while the blue region derives from a shape-fit to the  $p_{T,h}$ distribution.}
\end{center}
\end{figure}

On the left in Figure~\ref{Fig2}, we show the $p_{T,h}$ distribution in  $Wh$ production at $\sqrt{s} = 13 \, {\rm TeV}$ normalised to the SM prediction for  three different sets of Wilson coefficients.\hspace{0.5mm}\cite{Betalinprep}  In the case of $\bar c_{HW} = 0.03$~(blue), one observes a sizeable enhancement in the tail of the $p_{T,h}$ spectrum that amounts to around $50\%$ at $p_{T,h} \simeq 150 \, {\rm GeV}$, while for the choice $\bar c_{HW} = -\bar c_W = 2 \bar c_B = 0.03$~(orange) the event rate is reduced by about $-15\%$ with respect to the SM, almost independently of the precise $p_{T,h}$ value. The qualitative different behaviour of the two $p_{T,h}$ distributions can be understood by noticing that the leading $p_{T,h}^2$ dependence of $d\sigma/dp_{T,h}$ is proportional to the combination $\bar c_{HW} + \bar c_W$ of Wilson coefficients which is non-zero for the former but zero for the latter choice. In the case of $\bar c_6 = 10$ (red), one finally sees that the deviations in the $p_{T,h}$ spectrum change approximately linearly with $p_{T,h}$ and reach roughly $-10\%$ at $p_{T,h} \simeq 150 \, {\rm GeV}$. 

The observed shape differences can be used to better constrain the above benchmark cases compared to a fit that employs the information on the corresponding inclusive measurement  only. This feature is illustrated in the right panel of Figure~\ref{Fig2}, which displays the $95\%$~CL regions in the $\bar c_6\hspace{0.5mm}$--$\hspace{0.5mm} \bar c_{HW}$ plane that follow from a hypothetical HL-LHC measurement of the total cross section~(red) and the $p_{T,h}$ spectrum (blue) in the $Wh$ channel.\hspace{0.5mm}\cite{Betalinprep} From the plot it is evident that the fit to the inclusive measurement has a flat direction that allows for large correlated effects in $\bar c_6$ and $\bar c_{HW}$, while this degeneracy is resolved by the shape-fit to the  $p_{T,h}$ distribution. This simple example  illustrates nicely that differential Higgs measurements can give  important additional informations compared to standard Higgs-coupling fits. 

\section{Final words}

The overarching goal of this presentation was to emphasise that LHC precision measurements of rates and distributions in single-Higgs production can help to better constrain some of the Higgs interactions that are crudely known at present. The two  examples that we have discussed in some detail were the  charm Yukawa   and the Higgs trilinear coupling. In both cases it is important to stress that  to fully exploit the physics potential of the LHC one should try to combine all known search strategies. For the charm Yukawa coupling these are $h \to J/\psi \gamma$, $h \to c \bar c \gamma$, $pp \to Vc\bar c$, $pp \to hc$ as well as single-Higgs distributions, while for what concerns the trilinear Higgs coupling a combination of the constraints arising from $pp \to hh$, $pp \to h$ and the EW precision measurements seems essential.  The importance of measurements of distributions in gluon-fusion Higgs, $Vh$, VBF and $t \bar th$ production cannot be overemphasised in this context, since differential information has been shown to greatly enhance the sensitivity to the structure of the underlying theory.  Such measurements should therefore be pursued with vigour by both the ATLAS and CMS collaborations in the future LHC runs. 

\section*{Acknowledgments}

A  big thank you to Fady~Bishara, Wojtek Bizon, Martin Gorbahn, Pier Francesco Monni, Emanuele Re and Giulia Zanderighi for enjoyable collaborations on the topics discussed in this proceedings. I wish to thank the organisers of Recontres Moriond  EW 2017 for the invitation to give this  talk and I am grateful to many participant, in particular to Andre~David, Stefano~Di~Vita, Giuliano~Panico and Gavin~Salam, for interesting discussions. The continued hospitality and support of the CERN Theoretical Physics Department is highly appreciated.

\end{document}